\documentclass[prl,twocolumn,superscriptaddress,amsmath,amssymb,showpacs,floatfix,preprintnumbers]{revtex4}
\usepackage{amsmath}
\usepackage{graphicx}
\usepackage{dcolumn}
\usepackage{color}
\DeclareGraphicsExtensions{.png .jpg .pdf}

\begin{document}

\title{Efficient domain wall motion in asymmetric magnetic tunnel junctions with vertical current flow}

\author{S. Liu}\email{liux3824@umn.edu}
\affiliation{Department of Electrical and Computer Engineering, University of Minnesota, Minneapolis, Minnesota 55455, USA}
\author{D. J. P. de Sousa}\email{sousa020@umn.edu}\affiliation{Department of Electrical and Computer Engineering, University of Minnesota, Minneapolis, Minnesota 55455, USA}
\author{M. Sammon}\email{sammo017@umn.edu}\affiliation{Department of Electrical and Computer Engineering, University of Minnesota, Minneapolis, Minnesota 55455, USA}
\author{J. P. Wang}\email{jpwang@umn.edu}\affiliation{Department of Electrical and Computer Engineering, University of Minnesota, Minneapolis, Minnesota 55455, USA}
\author{T. Low}\email{tlow@umn.edu}
\affiliation{Department of Electrical and Computer Engineering, University of Minnesota, Minneapolis, Minnesota 55455, USA}

\date{ \today }

\begin{abstract}
In this paper, we study the domain wall motion induced by vertical current flow in asymmetric magnetic tunnel junctions. The domain wall motion in the free layer is mainly dictated by the current-induced field-like torque acting on it. We show that as we increase the MTJ asymmetry, by considering dissimilar ferromagnetic contacts, a linear-in-voltage field-like torque behavior is accompanied by an enhancement in the domain wall displacement efficiency and a higher degree of bidirectional propagation. Our analysis is based on a combination of a quantum transport model and magnetization dynamics as described by the Landau-Lifshitz-Gilbert equation, along with comparison to the intrinsic characteristics of a benchmark in-plane current injection domain wall device.

\end{abstract}


\maketitle

\section{Introduction}

\begin{figure}[t]
\centerline{\includegraphics[width = \linewidth]{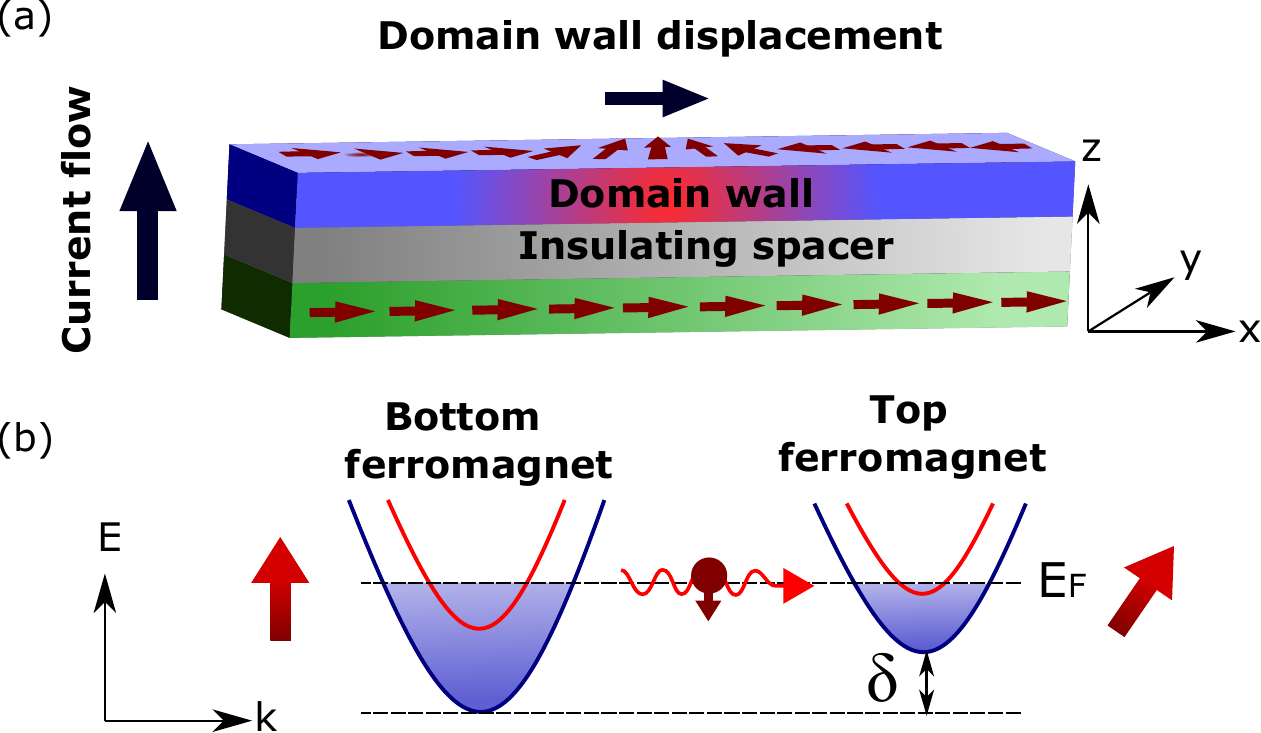}}
\caption{(Color online) \textbf{Domain wall (DW) device with vertical current flow}. (a) We consider a magnetic tunnel junction where an in-plane DW displacement in the free layer is induced by a vertical current flow. The DW on the top layer is represented by shaded red region. In this picture, tunneling electrons transfer spin angular momentum to the misaligned moments in the free layer via spin transfer torque, giving rise to a DW displacement. Top (free) and bottom (fixed) layer are assumed to be dissimilar ferromagnets, as represented by blue and green layers. (b) sketch of the top and bottom layer bands. Red and blue bands represent minority and majority spin bands, respectively, while the blue shaded regions represent the filled states for each ferromagnetic layer. The band filling of the bottom layer differs from the top one by a quantity $\delta$, which is the asymmetry parameter.}
\label{Fig1}
\end{figure}

Control of magnetic domain wall (DW) motion in nanowire-like structures using current injection has been of interest for memory applications due to promising qualities of low power consumption and non-volatility \cite{ref1}. The typical mode of operation for DW devices is current injection in-plane (CIP) along the propagation direction of the DW \cite{beach,chiba,grollier,incorvia}. However, current injection perpendicular to the plane (CPP) has the potential to be more space-efficient since it allows vertical integration of device structures. More importantly, CPP DW propagation has shown DW velocities up to 100 times faster than that of a CIP configuration at similar current densities \cite{ref2,ref3,boone}. This CPP configuration has the form of a perpendicular magnetic tunnel junction (MTJ), consisting of fixed and free magnetization layers that sandwich a MgO insulating layer shown in Fig. \ref{Fig1}(a). 

Current-induced propagation of the DW in the free layer of the MTJ relies on spin transfer torque (STT), where spin-polarized electrons interact with the misaligned magnetization of the DW to induce movement. The contribution of STT to magnetization dynamics in in-plane magnetized MTJs can be decomposed into two terms: a torque acting in-plane that will be referred to as the damping-like (DL) torque and an out-of-plane torque which is also known as the field-like (FL) torque. In such systems, it was demonstrated that the FL torque is the main contributor to DW motion \cite{ref3,boone}. This is due to the fact that in in-plane magnetized thin films, magnetic charges at top and bottom surfaces act to suppress the DL torque contribution as it tends to rotate the magnetization out-of-plane, leaving only the FL torque-induced DW displacements \cite{ref3,boone}. In addition, the voltage characteristics of these current-induced torques and their relative magnitudes were shown to depend on the degree of asymmetry of MTJ devices having dissimilar ferromagnetic contacts \cite{ref4, ref5, ref6}. Most notably, the asymmetry-induced tunability of the voltage dependence of the FL torque has important energy efficiency consequences when considering DW motion in the free layer in the CPP configuration, which has not been explored to-date.

In this work, we demonstrate that by increasing the asymmetry between the fixed and free layers of a MTJ, the voltage dependence of the FL torque is modulated, which leads to significantly improved current density and energy efficiency of DW propagation in the free layer. We also demonstrate that unlike symmetric MTJs, bidirectional DW motion can be achieved in asymmetric MTJs, allowing for new switching modality in these devices. Our analysis suggests that the faster DW velocities, in conjunction with improvements in energy efficiency with asymmetric MTJs can allow a CPP device to outperform counterpart CIP DW devices.

\section{Theoretical model}
 The system is sketch in Fig.~\ref{Fig1}(a). A top ferromagnetic layer with a DW, represented by the red shaded region, is separated from the bottom magnetic layer (fixed layer) with a single domain along the x direction by a thin insulating spacer. We assume that the top and bottom ferromagnets have in-plane magnetization with different band fillings as sketched by the blue shaded regions in the band diagrams of Fig.~\ref{Fig1}(b). The MTJ asymmetry is controlled by the band filling difference via the parameter $\delta$. 
In this section, we present the theoretical model employed in the description of DW motion due to current-induced torque mediated by the spin polarized tunneling electrons. A description of the quantum transport model used for investigating the evolution of spin transfer torques with the MTJ asymmetry is followed by the analysis of the Landau-Lifshitz-Gilbert (LLG) equation applied to the DW motion in the free layer.

\begin{figure*}[t]
\centerline{\includegraphics[scale = 1.4]{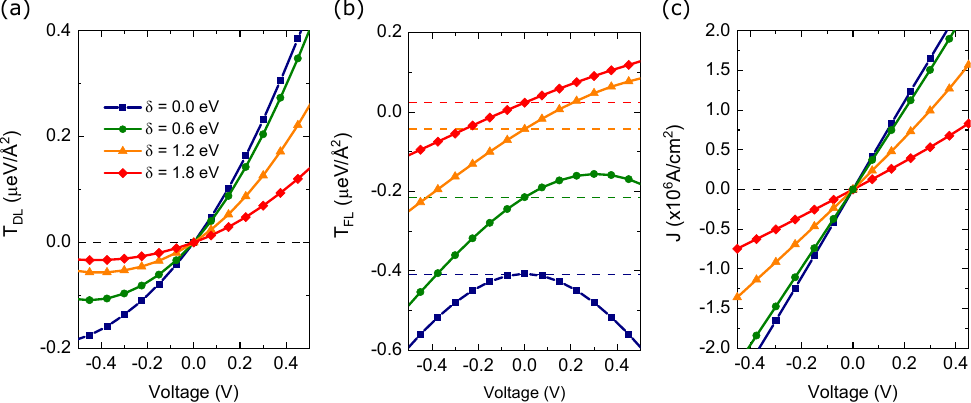}}
\caption{(Color online) \textbf{Voltage dependence of current-induced torques and current density}. Voltage dependence of the (a) DL, (b) FL current-induced torques and (c) current density for asymmetric MTJs for several asymmetry parameters $\delta$. The horizontal dashed lines highlight the equilibrium value of these quantities.}
\label{Fig2}
\end{figure*} 

\subsection{Current-induced torque}
Following Refs.~\cite{ref4, ref5, ref7, ref8, Sayeef}, we describe the non-equilibrium properties of the MTJ with a single-orbital tight-binding (TB) model in combination with quantum transport simulations. The system is composed of two semi-infinite magnetic leads coupled to an insulating spacer via spin-independent hopping matrices $H_{I,T(B)} = t\sigma_0$, where $\sigma_0 = I_{2\times 2}$ is the 2$\times$2 identity matrix in spin space and $t$ is the associated hopping parameter. The Hamiltonian of the coupled system has the block tridiagonal form
\begin{eqnarray}
H = \left(
\begin{array}{ccc}
H_T & H_{T,I} & 0 \\
H_{T,I}^{\dagger} & H_{I} & H_{I,B} \\
0 & H_{I,B}^{\dagger} & H_{B} \\
\end{array}
\right),
\label{eq1}
\end{eqnarray}
Where $H_{\alpha}$, with $\alpha = \rm{T},\rm{I},\rm{B}$, refers to the Hamiltonian block describing the top ferromagnet (T), the insulating spacer (I) and the bottom ferromagnet (B), respectively. Each region is assumed to be composed of several atomic planes and is described by the TB Hamiltonian
\begin{eqnarray}
H_{\alpha} = \sum_{ij,\sigma \sigma'} H_{ij,\alpha}^{\sigma\sigma'}c_{i\sigma}^{\dagger} c_{j\sigma'},
\label{eq2}
\end{eqnarray}
where the sum over atomic plane indexes $i,j$ extends over nearest-neighbors only and $c_{i\sigma}^{\dagger}$ ($c_{i\sigma}$) creates (annihilates) an electron with spin $\sigma$ in the $i$-th atomic plane. Each atomic monolayer is assumed to be translationally invariant along the in-plane directions. For simplicity, we assume that all hopping parameters are spin-independent with the same value $t = -1$ eV in all regions~\cite{ref4, ref5, ref6, ref7, ref8}. With this assumption, the spin dependent block Hamiltonian elements for each layer is given in terms of the spin-dependent onsite energy $\epsilon_{\sigma}(\textbf{k}_{||})$, where $\textbf{k}_{||}$ refers to the in-plane momenta, and nearest-neighbor hopping parameter $t$ as $H_{ij}^{\sigma\sigma'} = \delta_{\sigma\sigma'}\delta_{ij}\epsilon_{\sigma}(\textbf{k}_{||}) + \delta_{\sigma\sigma'} (\delta_{i, j+1} + \delta_{i, j - 1})t$ where $\sigma = \uparrow, \downarrow$ is the spin index and $\delta_{ij (\sigma\sigma')}$ is the Kronecker delta for real (spin) space labels. Assuming the interface is oriented along the $xy$ plane, the translation invariance along the in-plane directions allows us to express $\epsilon_{\sigma}(\textbf{k}_{||}) = \epsilon_{\sigma} + 2t(\cos(k_x) + \cos(k_y))$.

For the calculations presented in this paper, we assume $\epsilon_{\uparrow} - E_{\rm{F}} = 3.0$ eV, $\epsilon_{\downarrow} - E_{\rm{F}} = 5.6$ eV for the T ferromagnetic layer (free layer) with spin quantization axis along the $\textbf{x}'$ direction and $\epsilon_{\uparrow} - E_{\rm{F}} = 3.0 + \delta$ eV, $\epsilon_{\downarrow} - E_{\rm{F}} = 5.6 + \delta$ eV for the B ferromagnetic layer (fixed layer) with spin quantization axis along the $\textbf{x}$ direction, where the Fermi level is fixed at $E_{\rm{F}} = 0$ eV and the parameter $\delta$ controls the asymmetry between the magnetic layers~\cite{ref4, ref5, ref6}. Additionally, the tunnel barrier is assumed to be composed of $N = 3$ atomic planes with onsite energies $\epsilon_{\uparrow} = \epsilon_{\downarrow} = 9.0$ eV and the hopping parameters assume the values $t_{\uparrow} = t_{\downarrow} = -1.0$ eV. The particular choice of parameters provides a good estimate of the right order of magnitude of spin torques, exchange coupling and tunneling magnetoresistance (TMR) of Fe/MgO/Fe MTJs \cite{ref9, ref10, ref11}. 

In steady state, the spin transfer torque acting on the local moments at the $i$-th atomic plane of the top lead is $\textbf{T}_{i} = -[\nabla \cdot \textbf{Q}]_i$, where $[\nabla \cdot \textbf{Q}]_i = \textbf{Q}_{i-1,i} - \textbf{Q}_{i,i + 1}$ is the divergence of the spin current where
\begin{eqnarray}
\textbf{Q}_{ij} = \frac{1}{4\pi} \int_{BZ} \frac{d \textbf{k}_{||}}{(2\pi)^2}\int dE\operatorname{Tr}_{\sigma}[(H_{ji}G_{ij}^{<}-H_{ij}G_{ji}^{<})\vec{\sigma}],
\label{eq3}
\end{eqnarray} 
stands for the spin-current density between the $i$-th and $j$-th atomic planes. Additionally, $\vec{\sigma} = (\sigma_x, \sigma_y, \sigma_z)$ is a vector of Pauli matrices, $H_{ij}$ is hopping matrix between sites $i$ and $j$, $G_{ij}^{<}$ is the lesser Green function of the whole coupled system and the $\textbf{k}_{||}$ integration is performed over the 2D in-plane Brillouin zone (BZ). The total spin transfer torque acting on the top magnetic lead is $\textbf{T} = \sum_{i \in R} \textbf{T}_{i} =  \textbf{Q}_{0,1}$, corresponding to the spin current density between the last atomic layer of the insulating spacer ($0$-th layer) and the first atomic layer of the top magnetic layer ($1$-st layer). It is important to mention that the angular dependence of current-induced torques in MTJ devices is dictated only by the sine of relative angle, $\theta$, between the magnetization of the two ferromagnets \cite{ref4, ref5, ref6, ref7, ref8, ref9}. Thus, we only need to evaluate the voltage dependence of these torques for a single angular configuration, say $\theta = \pi/2$, in order to obtain their dependence over all important parameters for our analysis. When the system is driven out of equilibrium under an applied voltage $\mu_{\rm{T}} - \mu_{\rm{B}} = eV$ we assume that the potential drops linearly inside the oxide layer. 

Figure~\ref{Fig2} displays the voltage dependence of DL and FL torques, as well as the charge current density for several different values of the asymmetry parameter $\delta$. The horizontal dashed lines highlight the equilibrium values (at $\mu_T = \mu_B$) of these quantities. As is apparent from the blue symbols in Fig.~\ref{Fig2}(a), the current-induced DL torque exerted on the free layer of a symmetric MTJ ($\delta = 0$ eV) switches sign as one reverses the voltage polarity, being more efficient when spin polarized electrons flow from the fixed to the free layer, i.e., at $V>0$ according to our convention. Such feature enables one to achieve bipolar-bidirectional magnetization switching whenever the torque intensity surpasses the intrinsic damping of a free layer \cite{ref12, ref13}. The same qualitative features are also observed in asymmetric MTJs ($\delta \neq 0$ eV cases), but now with a smaller voltage modulation, i.e., smaller torque strength at a given voltage. It is worth emphasizing that the voltage modulation of the current density also drops with increasing $\delta$ such that the torque efficiency, defined as $T_{DL}/J$, might actually increase within some asymmetry window where the current density drops faster \cite{ref6}. 

The voltage dependence of the current-induced FL torque acting on the free layer is displayed in Fig.~\ref{Fig2}(b) for several asymmetry parameters $\delta$. First, one notices that the interlayer exchange coupling (IEC), being the magnitude of this torque at equilibrium, i.e., at $V=0$ V, changes as one tune the asymmetry of the MTJ. This feature, that is apparent by the shift in the horizontal dashed lines, can be understood in terms of Bruno's model for IEC \cite{ref14}, where its strength depends on the relative value of the reflection coefficients of majority and minority spins at both oxide/ferromagnet interfaces. Hence, as $\delta$ increases, electrons scatter differently as they encounter a larger energy barrier at the B interface, modifying the strength of the IEC between fixed and free layer mediated by the insulating spacer. Secondly, while the DL torque always presents qualitatively similar voltage dependencies with varying $\delta$, the FL torque might completely change its voltage dependence from purely quadratic ($\delta = 0$ eV) to linear (see, for example, the $\delta = 1.8$ eV case). Such features were predicted in the context of a similar TB approach and used to explain the FL torque voltage dependencies observed by different experimental groups \cite{ref4, ref5}. Such linear-in-voltage behavior of the FL torque was also predicted to be useful in assisting bipolar magnetization switching in asymmetric MTJs having non-negligible voltage control of magnetic anisotropy (VCMA) in the free layer \cite{ref6}.

In the next section, we explain how we combined the quantum transport results for the current-induced torques with the Landau-Lifshitz-Gilbert equation for describing DW motion in the free layer in the CPP configuration.

\subsection{Landau-Lifshitz-Gilbert equation}

The magnetization dynamics are described by the LLG equation with an added spin transfer torque term
\begin{eqnarray}
\frac{d\mathbf{m}}{dt} = -\gamma \mathbf{m}\times\mathbf{H}_{\textrm{eff}}+\alpha\mathbf{m}\times\frac{d\mathbf{m}}{dt}+\frac{\gamma}{\mu_0 M_S t_{\textrm{free}}}\mathbf{T}_{STT},
\label{llg}
\end{eqnarray}
where $\mathbf{m}$ is the magnetization of the free layer normalized with saturation magnetization $M_S$, $\gamma$ is the gyromagnetic ratio, $\alpha$ is the Gilbert damping parameter, $\mu_0$ is the vacuum permeability, and $t_{\textrm{free}}$ is the thickness of the free layer. $\mathbf{H}_{\textrm{eff}}$ is the effective magnetic field due to magnetostatic, exchange, and anisotropy energies added to any external magnetic field.

The total torque density due to STT is described as
\begin{eqnarray}
\mathbf{T}_{STT} = T_{FL}\mathbf{m}\times \mathbf{m}_p+T_{DL}\mathbf{m}\times(\mathbf{m}\times \mathbf{m}_p),
\label{torque}
\end{eqnarray}
where $\mathbf{m}$ ($\mathbf{m}_p$) is the magnetization direction of the free (pinned) layer, $T_{DL}$ is the magnitude of DL torque, and $T_{FL}$ is the magnitude of FL torque. The FL torque is also understood as the non-equilibrium interlayer exchange coupling. 

With the results of $T_{DL}$ and $T_{FL}$ from the quantum transport calculation in the previous section, the magnetization dynamics are solved using MuMax3, a GPU-accelerated micromagnetics simulator \cite{mumax}. The system simulated in MuMax3 is a Fe/MgO/Fe rectangular nanowire where the dimensions of the free layer are 15 nm wide and 3 nm thick. To evaluate the velocity of the DW, the length of the system is treated as infinite. The magnetic parameters used are of Fe, with $M_S=1.7\times10^6$ A/m, exchange stiffness $A_{\textrm{ex}}=1.3\times10^{-11}$ J/m, and Gilbert damping constant $\alpha=0.006$. The magnetization of the fixed layer of the MTJ is fixed in the +x direction. The magnetic system as described by Eq.~(\ref{llg}) is numerically solved on a 3-dimensional mesh where the mesh size is $1.5\times1.5\times1.5$ nm$^3$. Current is injected in the +z direction and is uniformly distributed across the wire cross-section. The calculation assumes zero temperature. The non-equilibrium torque density contributions as a function of voltage from the quantum transport calculation are interpolated to be used in the micromagnetics simulation.

\section{Results}
\begin{figure}[t]
\centerline{\includegraphics[width = \linewidth]{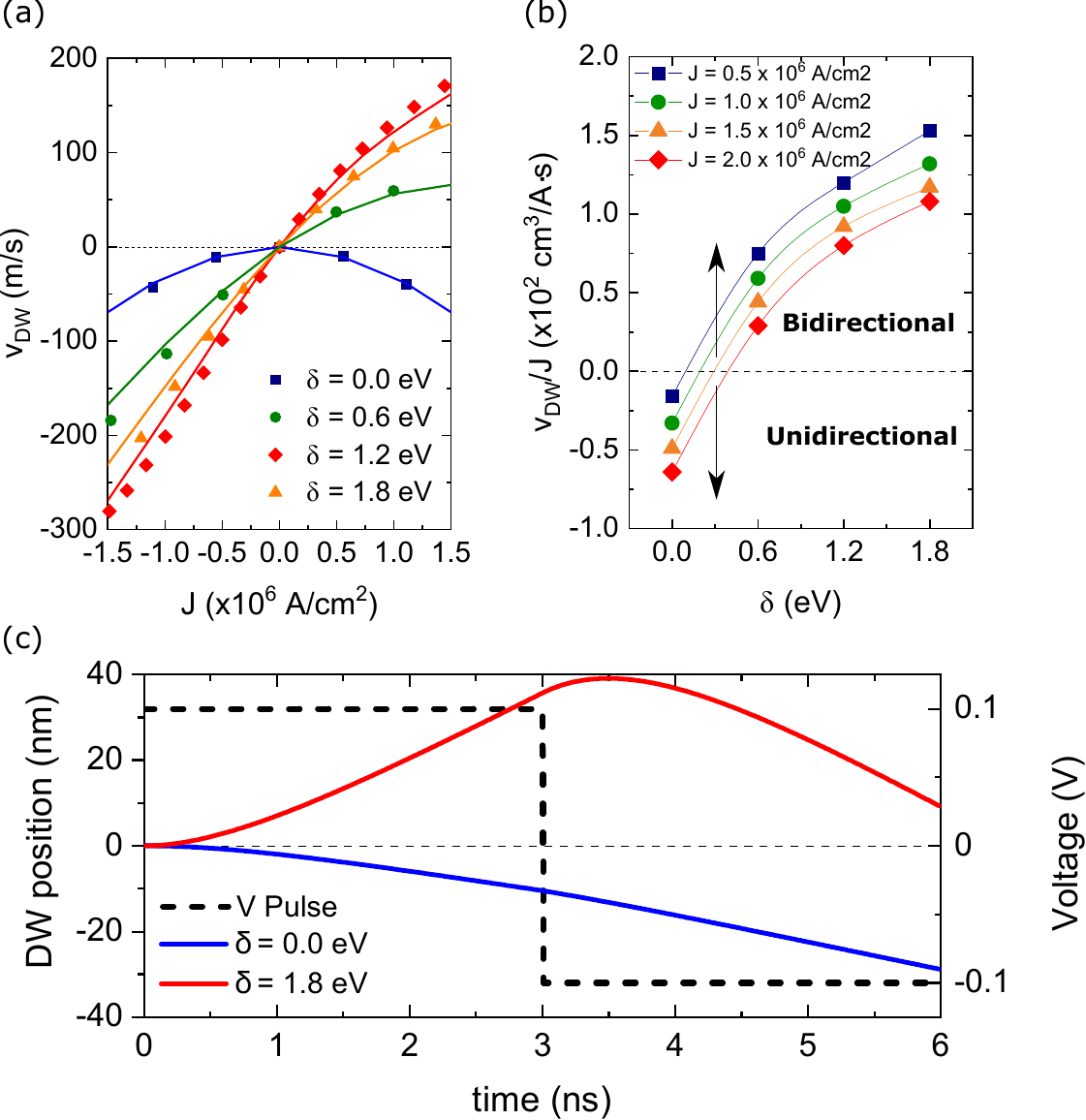}}
\caption{(Color online) \textbf{DW velocity characteristics for asymmetric MTJs}. (a) DW velocity as a function of the average current density for different asymmetries $\delta$s. The symbols were obtained numerically by solving the LLG equation for the rectangular nanowire with current-induced torques obtained from the quantum transport simulations. The solid lines are the analytical toy model results. (b) Asymmetry-dependent DW velocity per applied current density. Different symbols/curves were obtained by considering different applied current densities. Unidirectional and bidirectional highlighted regions for negative and positive $V_{\textrm{DW}}/J$, respectively, refers to the directionality of DW motion for each asymmetry case. (c) DW position for two cases $\delta = 0.0$ eV and $\delta = 1.8$ eV as a function of time. The voltage pulse is also shown as a solid black curve. }
\label{Fig3}
\end{figure} 

In this section, we present the DW propagation results along with analysis using a 1-dimensional model. We also present intrinsic delay and energy efficiency benchmarks of a proposed vertical current injection logic device and discuss the advantages of a MTJ structure.

\subsection{Domain wall velocity}

To produce the micromagnetics results of DW velocity dependence on current density, a constant voltage was applied perpendicularly across the MTJ where a positive voltage leads to electron flow from the fixed layer to the free layer. The position of the DW as a function of time was tracked using the average +x-direction magnetization of the nanowire and the steady-state velocity of the DW was recorded. This was repeated multiple times, changing the voltage for each trial. Using the average current density per voltage shown in Fig.~\ref{Fig2}, the dependence of DW velocity on current density can be obtained. 

It can be seen in Fig.~\ref{Fig3}(a) that for the symmetric case, DW velocity is characteristically quadratic, leading to unidirectional motion. With increasing asymmetry, the dependence of DW velocity on current density becomes characteristically linear which leads to bidirectional DW motion in the small current density regime. The change from even to odd dependence in voltage is reflective of the quadratic and linear FL torque dependence with voltage in symmetric and asymmetric MTJs, respectively. Additionally, the results indicate that a velocity of over 100 m/s can be obtained with current density 1$\times10^6$ A/cm$^2$ for a more asymmetric MTJ, which is more efficient than that of in-plane current injection \cite{ref1}.

A 1-dimensional analytical approximation described by Khvalkovskiy et al. can be used to characterize the DW motion \cite{ref3}:
\begin{subequations}
\begin{eqnarray}
\frac{u}{\Delta}-\alpha\dot{\Phi} - \gamma\frac{T_{DL}}{M_St_{\textrm{free}}}=\frac{\gamma}{M_S}K_{\textrm{x}}\sin(2\Phi), \\
-\alpha\frac{u}{\Delta}-\dot{\Phi} + \gamma\frac{T_{FL}}{M_St_{\textrm{free}}} = 0.
\label{analyt}
\end{eqnarray}
\end{subequations}

Here, $u$ is the DW velocity, $\Delta$ is the width of the DW, $K_{\textrm{x}}$ is the anisotropy constant in the x-direction, and $\Phi$ is the precessional angle of magnetization of the DW. 
\begin{eqnarray}
u = \frac{\gamma T_{FL}}{\alpha M_St_{\textrm{free}}}\Delta
\label{soln}
\end{eqnarray}
The steady state solution of the analytical approximation is $\dot{\Phi} = 0$, which yields the expression shown in Eq.~(\ref{soln}). This approximation indicates that DW velocity in this configuration is solely dependent on FL torque. Utilizing DW width $\Delta$ as the fitting parameter, $\Delta=11.7$ nm was obtained for minimum mean squared error while the actual DW width calculated from the magnetization profile is $\Delta=12.1$ nm. From this analysis, the analytical model is in good agreement with the micromagnetics data.

In order to quantify the efficiency of DW propagation, the figure-of-merit of DW velocity divided by current density is presented. The results for micromagnetics simulation are shown in Fig.~\ref{Fig3}(b). The results here indicate a clear trend; for small current densities, asymmetry significantly increases the efficiency of DW propagation in terms of current density. The increase in efficiency can be traced to the fact that increased asymmetry leads to more dominant field like torque contributions, which plays a more critical role in DW motion in this geometry. Additionally, the figure-of-merit for the symmetric MTJ ($\delta = 0$ eV) is negative, indicating unidirectional DW propagation. This is in contrast to the case of an asymmetric MTJ, where bidirecitonal DW propagation allows for greater control of DW dynamics. 

An important point to emphasize is the bidirectionality of the DW motion at different asymmetry levels. Figure~\ref{Fig3}(a) and (b) suggest that DW motion in symmetric MTJs, having $\delta = 0.0$ eV, is purely unidirectional, i.e., the vertical current flow induces movement along a particular direction for both voltage polarities, being incapable of reversing the DW motion to the opposite direction due to the quadratic voltage dependence of the FL torque in this case. Asymmetric MTJs, however, display a linear-in-voltage dependence of the FL torque. Therefore, asymmetric MTJ devices (with $\delta \neq 0$ eV) are more appropriate for applications were the bidirectionality of the DW movement via electrical means is required. To illustrate this difference, we plotted the DW position as a function of time for symmetric ($\delta = 0.0$ eV) and asymmetric ($\delta = 1.8$ eV, as an example) MTJ devices under a voltage pulse of absolute value of $0.1$ V that reverses sign at $t = 3$ ns, as shown in Fig.~\ref{Fig3}(c). Before $t = 3$ ns where the voltage is positive, DWs for symmetric and asymmetric devices move in opposite directions. After the voltage polarity reversal at $t = 3$ ns, one observes that DW motion of symmetric MTJ is not reversed while that of the asymmetric device is, i.e., asymmetric devices offer a greater flexibility when considering voltage control of DW motion under vertical current flow. We also observe that the DW in asymmetric MTJs propagates much farther than DW in symmetric devices, as is apparent in Fig.~\ref{Fig3}(c).

\subsection{Intrinsic device benchmarking}

In this section, a comparison in terms of energy dissipation is made between the presented CPP device and a benchmark CIP device. While most benchmarks require a device architecture beyond the scope of this paper, intrinsic benchmark comparisons can be made. The benchmark device is a 3-terminal MTJ device utilizing in-plane current injection \cite{incorvia}. Using the same methodology as in Nikonov et al., the intrinsic energy dissipation for a logic operation $E_{\textrm{int}}$ can be calculated as follows \cite{nikonov}:
\begin{eqnarray}
E_{\textrm{int}} = \int_0^{t_{\textrm{int}}}V(t)*[J(t)*L*W]dt
\label{device}
\end{eqnarray}
where $t_{\textrm{int}}$ is the time delay taken for the DW to propagate past the sensing MTJ, $V(t)$ and $I(t)$ are the voltage and current, and $L$ and $W$ are the length and width of the channel.

\begin{figure}[t]
\centerline{\includegraphics[width = \linewidth]{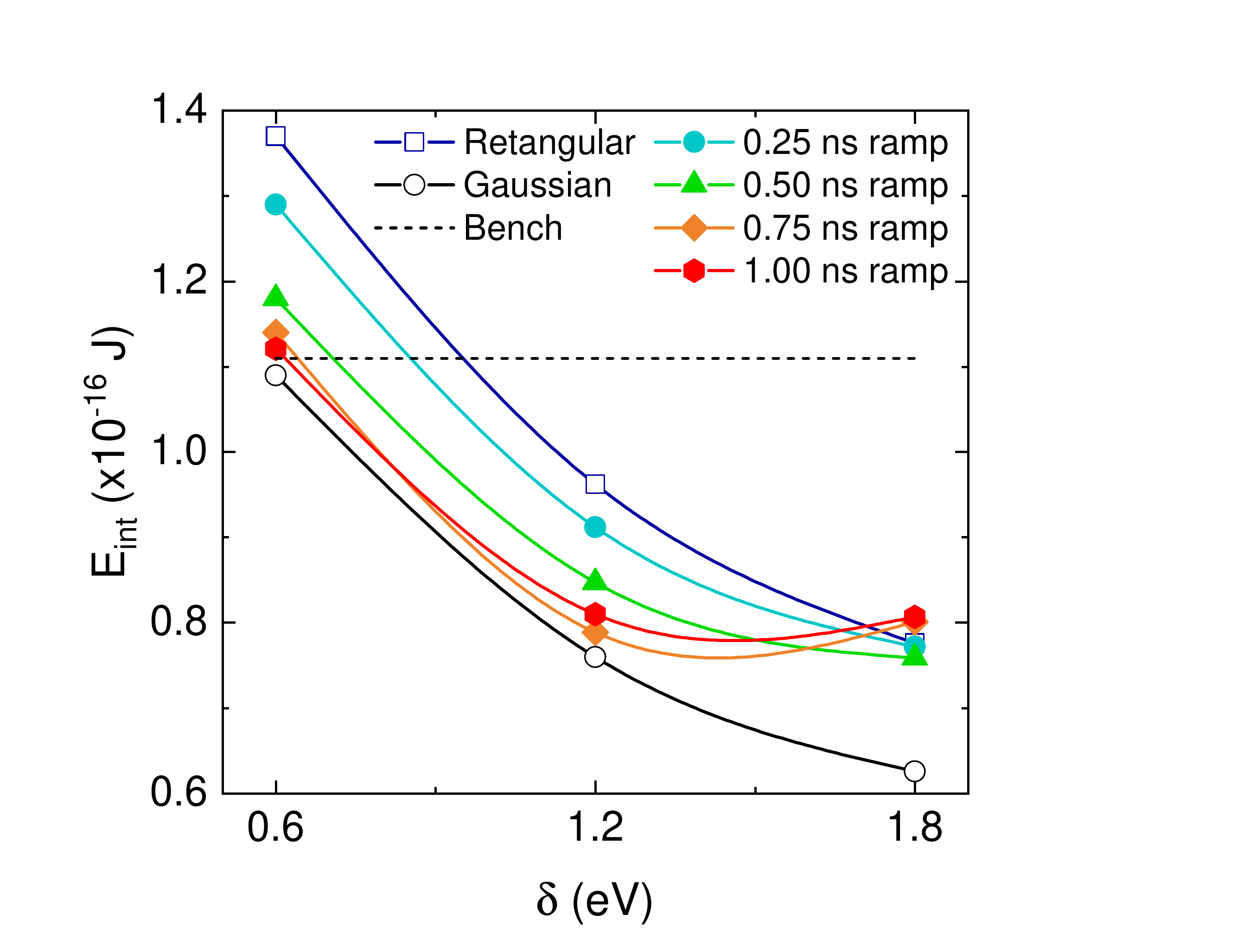}}
\caption{(Color online) \textbf{CPP vs. CIP}. CPP device intrinsic energy dissipation considering different voltage pulses: Rectangular (open squares), ramp with different rise times ranging from 0.25 ns to 1.00 ns (filled symbols) and Gaussian (open circle) shaped voltage pulses. The horizontal dashed line is the CIP benchmark from Ref.~\cite{nikonov}}
\label{Fig4}
\end{figure} 

In order to make comparisons to the in-plane current injection device, the time delay is set at $t_{\textrm{int}}=1.77$ ns. The width and thickness of the free layer are set as $W = 15$ nm and $t = 1.5$ nm respectively. The length of the sensing MTJ is set to be $30$ nm, same as that used in Nikonov et al. at twice the width of the channel. The length of the channel greatly affects the energy dissipation of the device, and here it is set to be 1.5 times the size of the sensing MTJ to accommodate extra length required for pinning sites at the ends of the nanowire. Additionally, for standardization purposes, the input waveform voltage amplitude is set to be 0.1 V. The input waveforms used are rectangular, variable timing ramps, and Gaussian. The DW dynamics are simulated using MuMax3 and the DW position is tracked using the method previously described in the DW velocity section. 

For the results in Fig.~\ref{Fig4}, the case for a symmetric MTJ ($\delta=0$ eV) is not included because it is not possible to obtain a time delay $t_{\textrm{int}}=1.77$ ns. For the degrees of asymmetry presented,  the energy dissipation is generally monotonically decreasing for increasing asymmetry, indicating increased device efficiency. However, for the 0.75 and 1.00 ns ramps, there is a slight increase in energy dissipation between $\delta=1.2$ eV and $\delta=1.8$ eV. Additionally, for $\delta=1.8$ eV, the effect seems to become more pronounced as the ramp time increases. This is likely the result of the nonlinear relationship between voltage, current density, and non-equilibrium FL torque. However, across the more experimentally accessible range of $\delta$, it can be concluded that device intrinsic efficiency increases as device asymmetry increases. Additionally, the efficiency of the proposed CPP device can outperform the CIP benchmark device (horizontal dashed line) with appropriate tuning of device asymmetry. It is also found that the Gaussian pulse (more adiabatic) results in the most efficient switching while the rectangular pulse (more abrupt) generally results in the least energy efficient switching. 

\section{Conclusions}

To summarize our results, we use a quantum transport model in conjunction with a micromagnetics model described by the LLG equation to demonstrate that DW propagation efficiency in a vertical current injection MTJ with in-plane magnetization can be significantly increased by increasing the asymmetry of the ferromagnetic contacts. This is due to the fact that DW propagation with this geometry is dominated by the FL torque; as the asymmetry of the MTJ is increased, the voltage dependence of the FL torque acting on the free layer is altered. We present two different methods to quantify DW propagation efficiency, with current density as well as energy dissipation of a proposed device that utilizes vertical current injection. In both cases, there is a clear general trend of increased efficiency with increased asymmetry. In particular, this increased asymmetry can significantly improve the performance of the proposed device and potentially outperform a CIP benchmark device. Due to the current efficiency limitations of vertical current injection, it may not be perfectly suitable to implement in a logic device. However, these results have promising implications for applications like spintronic interconnects, where vertical current injection has already been shown to be effective \cite{interconnect}. We show that by tuning the asymmetry of a vertical current injection magnetization in-plane MTJ, increased DW propagation efficiency as well as bidirectional DW motion can be achieved.

\textit{Acknowledgments}. DS, TL, and JPW were partially supported by DARPA ERI FRANC program. We acknowledge useful discussions with D. E. Nikonov from Intel Corporation.

\end{document}